\documentclass[12pt,epsf,epsfig,psfig]{article}
\usepackage{graphicx}
\usepackage{epsfig}
\def\J{$J/\psi$}
\def\j{J/\psi}
\def\X{$\chi_c$}
\def\x{\chi}
\def\P{$\psi'$}
\def\p{\psi'}
\def\U{$\Upsilon$}

\def\C{c{\bar c}}

\def\Q{Q{\bar Q}}
\def\e{\epsilon}

\def\l{\Lambda_{\rm QCD}}

\def\NP{{ Nucl.\ Phys.\ }}
\def\PL{{ Phys.\ Lett.\ }}
\def\PR{{ Phys.\ Rev.\ }}

\def\PRL{{ Phys.\ Rev.\ Lett.\ }}

\def\ZP{{ Z.\ Phys.\ }}
\def\EPJ{{Eur.\ Phys.\ J.\ }}

\def\be{\begin{equation}}
\def\ee{\end{equation}}
\def\lsim{\raise0.3ex\hbox{$<$\kern-0.75em\raise-1.1ex\hbox{$\sim$}}}
\def\gsim{\raise0.3ex\hbox{$>$\kern-0.75em\raise-1.1ex\hbox{$\sim$}}}

\begin{document}

\thispagestyle{empty}
September 19, 2006 \hfill BI-TP 2006/34

\vskip 2cm

\centerline{\Large \bf Quarkonium Binding and Dissociation:}

\bigskip

\centerline{\Large \bf The Spectral Analysis of the 
QGP*}

\vskip1cm

\centerline{\large \bf Helmut Satz} 

\bigskip

\centerline{\sl Fakult\"at f\"ur Physik, Universit\"at Bielefeld}

\centerline{\sl Postfach 100 131, D-33501 Bielefeld, Germany}
 
\vskip1cm

\centerline{\bf Abstract:}

\vskip0.5cm

In statistical QCD, the thermal properties of the quark-gluon plasma can be 
determined by studying the in-medium behaviour of heavy quark bound states. 
The results can be applied to quarkonium production in high energy nuclear 
collisions, if these indeed form a fully equilibrated QGP. Modifications 
could arise if an initial charm excess persists in the collision evolution 
and causes quarkonium regeneration at hadronization.

\vfill

\hrule width6cm height0.5pt\hfil\break

\vskip-0.4cm

{\large *} Invited talk at the {\sl 2$^{nd}$ International Conference on
Hard and Electromagnetic Probes of High Energy Collisions}, 
Asilomar/California, June 9 - 15, 2006, and at the {\sl International
Workshop on Heavy Quarkonium 2006}, Brookhaven National Laboratory,
June 27 - 30, 2006.

\newpage

\noindent{\large \bf 1.\ Introduction}

\bigskip

We know from statistical QCD that strongly interacting matter undergoes
a deconfinement transition to a new state, the quark-gluon plasma. How
can we study this state - which phenomena provide us with information 
about its thermal properties? The main probes considered so far are 
\begin{itemize}
\vspace*{-0.2cm}
\item{e-m signals (real or virtual photons)}
\vspace*{-0.2cm}
\item{heavy flavours and quarkonia ($\Q$ pairs)}
\vspace*{-0.2cm}
\item{jets (energetic partons)}
\end{itemize}
The ultimate aim must be to carry out {\sl ab initio} calculations
of the in-medium behaviour of these probes in finite temperature QCD.

\medskip

In high energy nuclear collisions, we want to study the deconfinement
transition and the QGP in the laboratory. The ultimate aim here must
be to show that experimental results confirm the predictions of
statistical QCD, or that they disagree with them. If the latter should
happen, we can unfortunately not conclude that statistical QCD is wrong;
a more likely conclusion would be that nuclear collisions do not produce
the medium studied in equilibrium QCD thermodynamics.  

\medskip

I want to consider here a specific case study for the program just outlined:
the spectral analysis of quarkonia in a hot QGP and its application to
nuclear collisions.

\medskip

The theoretical basis for this analysis is:
\begin{itemize}
\vspace*{-0.2cm}
\item{The QGP consists of deconfined colour charges, so that the binding
of a $\Q$ pair
is subject to the effects of colour screening.}
\vspace*{-0.2cm}
\item{The screening radius $r_D(T)$ decreases with temperature $T$.}
\vspace*{-0.2cm}
\item{When $r_D(T)$ falls below the binding radius $r_i$ of a $\Q$ state 
$i$, the $Q$ and the $\bar Q$ can no longer bind, so that quarkonium $i$ 
cannot exist \cite{MS}.}
\vspace*{-0.2cm}
\item{The quarkonium dissociation points $T_i$, specified through 
$r_D(T_i)\simeq r_i$, thus determine the temperature of the QGP, as 
schematically illustrated in Fig. \ref{spectral}.} 

\begin{figure}[htb]
\centerline{\epsfig{file=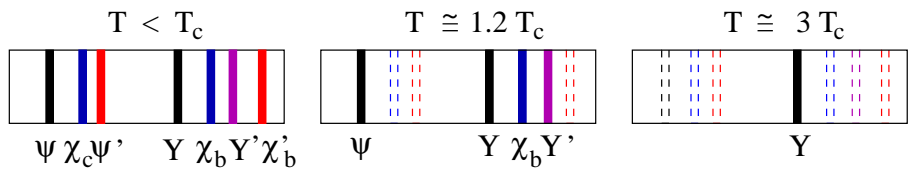,width=12cm}}
\caption{Quarkonium spectral lines as thermometer} 
\label{spectral}
\end{figure}

\end{itemize}

\begin{figure}[htb]
\centerline{\epsfig{file=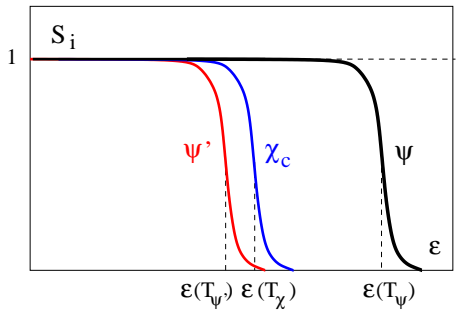,width=6cm}}
\caption{Charmonium survival probabilities vs.\ energy density} 
\label{survival}
\end{figure}

Experimentally, quarkonium studies also provide a great tool:
\begin{itemize}
\vspace*{-0.2cm}
\item{In $AA$ collisions, quarkonium production can be measured as function 
of collision energy, centrality, transverse momentum, and $A$.}
\vspace*{-0.2cm}
\item{The onset of (anomalous) suppression for the different quarkonium 
states can be determined and correlated to thermodynamic variables, such 
as the temperature or the energy density.}
\vspace*{-0.2cm}
\item{The resulting thresholds in the survival probabilities $S_i$ of 
states $i$ can then be compared to the relevant QCD predictions, as
illustrated in Fig.\ \ref{survival}.}
\end{itemize}

In this way we can, at least in principle, obtain a direct comparison
between experimental results and quantitative predictions from finite
temperature QCD.

\vskip0.5cm

\noindent{\large \bf 2.\ In$~\!$-Medium Behaviour of Quarkonia: Theory}

\bigskip

We consider as quarkonia those bound states of {\sl heavy} quarks which
are {\sl stable} under strong decay; they are thus pairs of charm 
($m_c \simeq 1.3$ GeV) or beauty ($m_b \simeq 4.7$ GeV) quarks, whose
overall masses fall below the open charm or beauty thresholds. The large
quark mass allows spectroscopy based on non-relativistic potential 
theory \cite{Schrodinger}. Hence the Schr\"odinger equation
\vskip-0.2cm
\be
\left\{2m_c -{1\over m_c}\nabla^2 + V(r)\right\} \Phi_i(r) = M_i \Phi_i(r),
\label{schroedinger}
\ee
\vskip0.1cm
using the ``Cornell'' form for the confining potential \cite{Cornell},
\be
V(r) = \sigma ~r - {\alpha \over r},
\ee 
in terms of the string tension  $\sigma \simeq 0.2$ GeV$^2$ and the 
gauge coupling $\alpha \simeq \pi/12$, determines the masses $M_i$ 
and the radii $r_i$ of the different charmonium and bottomonium states. 
The results are summarized in Table 1 and are seen to give a good
account of quarkonium spectroscopy, with an error of less than 1\% in
the mass determination $\Delta M$ for all (spin-averaged) states.

\vskip0.5cm

\hskip1.5cm
\renewcommand{\arraystretch}{1.6}
\begin{tabular}{|c|c|c|c|c|c|c|c|c|}
\hline
{\rm state}& $J/\psi$ & $\chi_c$ & $\psi'$  & $\Upsilon$
 & $\chi_b$ & 
$\Upsilon'$ & $\chi_b'$ & $\Upsilon''$ \\
\hline
{\rm mass~[GeV]}&
3.10&
3.53&
3.68&
9.46&
9.99&
10.02&
10.26&
10.36 \\
\hline
$\Delta E$ {\rm[GeV]}&0.64&0.20&0.05 &1.10&
0.67&0.54&0.31&0.20 \cr
\hline
$\Delta M$ {\rm[GeV]}&0.02&-0.03&0.03 & 0.06&
-0.06&-0.06&-0.08&-0.07 \cr
\hline
{\rm radius~[fm]}&0.25&0.36&0.45&
0.14&0.22& 0.28& 0.34 &0.39 \cr
\hline
\end{tabular}

\vskip0.6cm

\centerline{Table 1: Quarkonium spectroscopy in non-relativistic potential 
theory \cite{HS-jpg}}

\newpage

The charmonium and bottomonium ground states are thus tightly bound, with
a binding energy $\Delta E = 2M_{D,B} - M_0 \gg \l \simeq 0.2$ GeV, and 
very small, with $r_0 \ll r_h \simeq 1$ fm. What happens to them in a QGP?

\medskip

The effect of colour screening is that the binding becomes weaker and of
shorter range. When the force range or the screening radius fall below the
binding radius, the $Q$ and $\bar Q$ can no longer ``see'' each other, and
hence the bound state becomes dissociated. As already noted, the 
quarkonium dissociation points therefore determine the temperature and 
thus also the energy density of the QGP. The basic question thus is how
to calculate the quarkonium dissociation temperatures.

\medskip

Early attempts were based on models of the heavy quark potential, 
essentially obtained from $d=1$ electrodynamics. Using these in the
Schr\"odinger equation, together with a crude lattice form for the
screening mass, led to first predictions \cite{KMS,KS91}
\be
T_{\j}~\gsim~T_c,~~~ T_{\x}~\&~T_{\p}~\lsim~ T_c,
\ee
where $T_c$ is the deconfinement temperature.

\medskip

When lattice results for the heavy quark potential became available from
finite temperature lattice studies, these were employed in the Schr\"odinger
equation in various forms \cite{DPS1} - \cite{MP}. The results eventually
converged fairly well and the present status is schematically summarized 
in Table 2.
 
\vskip0.5cm

\centerline{
\renewcommand{\arraystretch}{1.5}
\begin{tabular}{|c||c|c|c|}
\hline
$\rm state$ & $J/\psi(1S)$ & $\chi_c(1P)$ & $\psi^\prime(2S)$\\
\hline
\hline
$T_d/T_c$ & $2.1$  & $1.2$ & $1.1$\\
\hline
\end{tabular}}

\vskip0.6cm

\centerline{Table 2: Charmonium dissociation temperatures in lattice-based
potential theory}

\bigskip

Both previous approaches assume the validity of a two-body potential 
treatment at finite temperature near a critical point. This assumption
is no longer necessary if the quarkonium spectrum can be calculated
directly in finite temperature lattice QCD. Such calculations have
become possible in recent years and results were presented by several
groups, first in quenched QCD \cite{Umeda} - \cite{Iida} and now also
in full (two-flavour) QCD \cite{Morrin,Aarts}. The present state of these
results is summarized in Table 3.

\vskip0.5cm

\centerline{
\renewcommand{\arraystretch}{1.5}
\begin{tabular}{|c||c|c|c|}
\hline
$\rm state$ & $J/\psi(1S)$ & $\chi_c(1P)$ & $\psi^\prime(2S)$\\
\hline
\hline
$T_d/T_c$ & $ > 2.0$  & $< 1.1$ & ?\\
\hline
\end{tabular}}

\vskip0.6cm

\centerline{Table 3: Charmonium dissociation temperatures from finite 
temperature} 

\centerline{Lattice QCD calculations}

\bigskip

Very recently also first lattice results have been presented for 
bottomonium dissociation in quenched QCD \cite{Datta-b,Velytsky}; one 
finds there
\be
T_{\Upsilon}~\gsim~2~T_c, ~~~T_{\x_b}~\lsim~1.15~T_c. 
\ee
The low value reported for the $\x_b$, which has approximately the same
binding energy as the \J, remains at present quite puzzling.

\medskip

We thus find from direct finite temperature lattice studies,
both in quenched and in full QCD, as well as in lattice-based potential
work, that the \J~and the \U~survive up to $T\geq 2~\!T_c$, which means up
to energy densities of 25 GeV/fm$^3$ or more. In contrast, the \X~and
(so far only from potential studies) the \P~melt near $T_c$, i.e., for
energy densities in the range 0.5 - 2.0 GeV/fm$^3$. It should be noted
that ``survival'' here means that the corresponding signal is seen
up to the temperature in question. So far, lattice QCD results do not
yet allow a determination of the widths as function of temperature, and
hence it is not known if even the ground states acquire a very large
width with increasing $T$. Moreover, the comparison of lattice and
potential theory can be carried out on a more detailed level than given 
by just the dissociation temperatures. A study of correlators in both
approaches can thus certainly provide more insight \cite{MP}.

\medskip

In closing this section, we summarize the modifications in theory which 
have led to the present new theoretical view, in particular of \J~survival
in a hot QGP: 
\begin{itemize}
\vspace*{-0.1cm}
\item{Earlier lattice studies had provided only the colour average of
the in-medium $\Q$ free energy; now it is possible to separate
out the colour singlet contribution.}
\vspace*{-0.2cm}
\item{Earlier potential models had used the free energy as potential
in the Schr\"odinger equation; today we can specify the colour singlet
internal energy, which is more realistic as the relevant potential
and leads to a stronger binding.} 
\vspace*{-0.2cm}
\item{There now exist direct finite temperature lattice QCD
studies of the in-medium behaviour of charmonia, allowing {\sl ab initio}
conclusions not based on any potential model, and they support a higher
\J~dissociation temperature.}
\end{itemize}
\vspace*{-0.1cm}
What then does this imply for quarkonium production as QGP probe 
in nuclear collisions?

\vskip0.5cm

\noindent{\large \bf 3.\ In$~\!$-Medium Behaviour of Quarkonia: Phenomenology}

\bigskip

The modifications observed when comparing \J~production in $AA$ collisions
to that in $pp$ interactions have two distinct origins. Of primary
interest is obviously the effect of the secondary medium produced in the
collision - this is the candidate for the QGP we want to study. In 
addition, however, the presence of cold nuclear matter in target and 
projectile can also affect the production process and final rates. This 
ambiguity in the origin of any observed \J~suppression thus has to be resolved.

\medskip 

A second empirical feature to be noted is that the measured \J~production 
consists of directly produced $1S$ states as well as of feed-down from 
\X(1P) and \P(2S) decay. In the previous section, we had seen that a hot 
QGP affects the higher excited quarkonium states much sooner 
(at lower temperatures) than the ground states. This results in another
ambiguity in observed \J~production - are only the higher excited states
affected, or do all states suffer?

\medskip

An ideal solution of these problems would be to measure separately
\J, \X~and \P~production first in $pA$ (or $dA$) collisions, to determine
the effects of cold nuclear matter, and then measure, again separately,
the production of the different states in $AA$ collisions as function of
centrality at different collision energies. While the production of the 
\P~has been studied in both $pA$ and $AA$ collisions, \X~data on nuclear
targets are not yet provided.

\medskip

Until such data become available, we resort to a more
operational approach, whose basic features are:
\begin{itemize}
\vspace*{-0.2cm}
\item{We assume that the \J~feed-down rates in $pA$ and $AA$ are the same 
as in $pp$, i.e., 60 \% direct $\j(1S)$, 30 \% decay of $\x(1P)$, 
and 10 \% decay of $\p(2S)$.}
\vspace*{-0.2cm}
\item{We specify the effects due to cold nuclear matter by a Glauber 
analysis of $pA$ or $dA$ experiments in terms of $\sigma_{abs}^i$ for 
$i=$\J, \X, \P. This $\sigma_{abs}^i$ is not meant as a real cross-section
for charmonium absorption by nucleons in the nucleus; it is rather used to
parametrize all initial and final state nuclear effects, including
shadowing/antishadowing, parton energy loss as well as pre-resonance
and resonance absorption.}
\vspace*{-0.2cm}
\item{In the analysis of $AA$ collisions, we then use $
\sigma_{abs}^i$ 
in a Glauber analysis to obtain the predicted form of {\sl normal} 
\J~suppression. This allows us to identify {\sl anomalous} \J~suppression
as the difference between the observed production distribution and that
expected from only normal suppression. We parametrize the anomalous
suppression through the survival probability
\be
S_i = {(dN_i/dy)_{\rm exp} \over (dN_i/dy)_{\rm Glauber}}
\ee
for each quarkonium state $i$.}
\end{itemize}

\medskip

\begin{figure}[htb]
\centerline{\epsfig{file=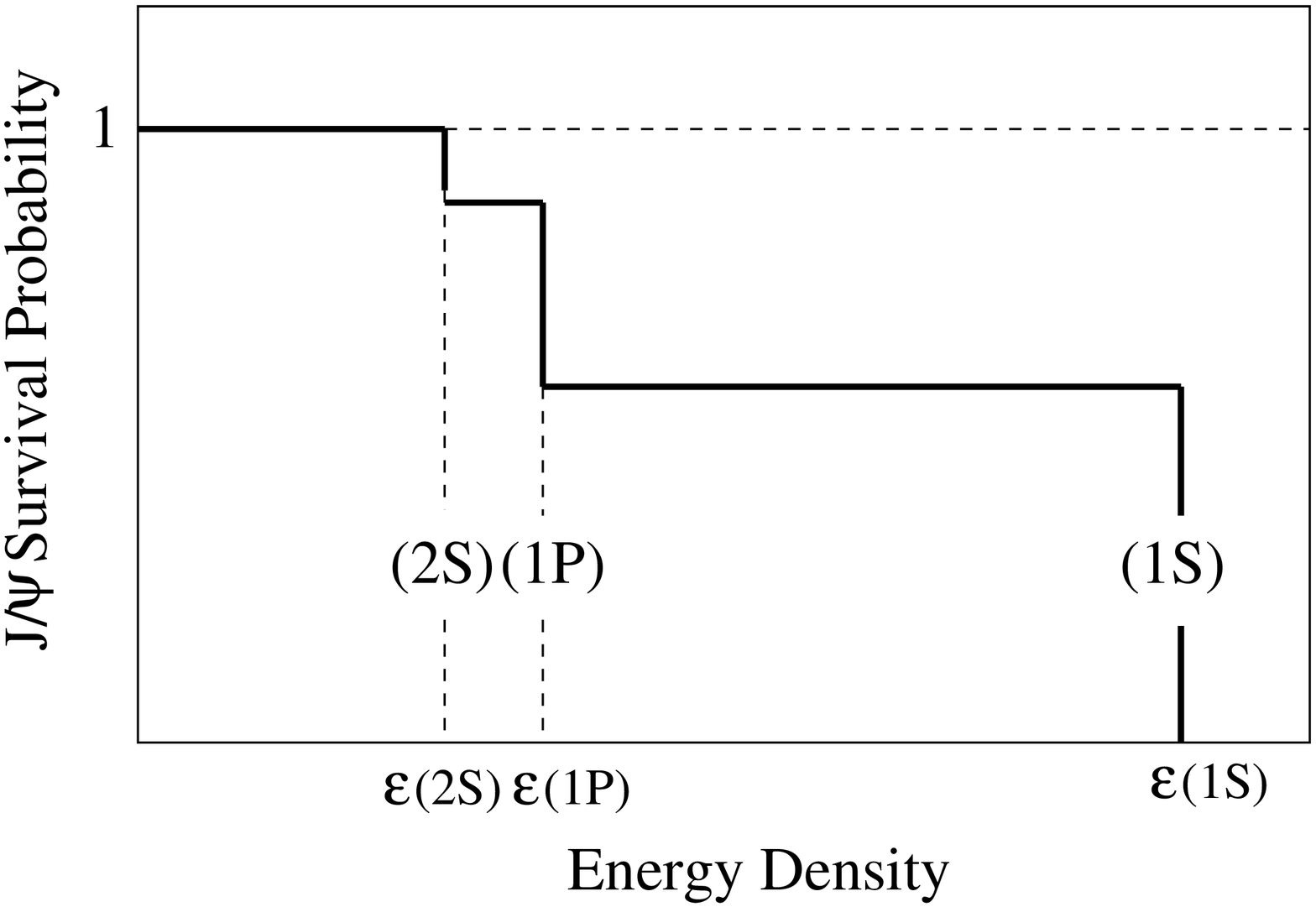,width=6.5cm}}
\caption{Sequential \J~suppresssion}
\label{sequential}
\end{figure}

With the effects of cold nuclear matter thus accounted for, what form do
we expect for anomalous \J~suppression? If $AA$ collisions indeed produce
a fully equlibrated QGP, we should observe a sequential suppression pattern 
dfor \J~and \U, with thresholds predicted (in terms of temperature or
energy density) by finite temperature QCD \cite{KMS,KS91}, 
\cite{G-S}-\cite{KKS}. The resulting pattern for the \J~is illustrated in 
Fig.\ \ref{sequential}.

\medskip

Its consequences are quite clear. If, as present statistical QCD studies
indicate, the direct $\j(1S)$ survives up to about $2~T_c$ and hence to
$\e \geq 25$ GeV/fm$^3$, then all anomalous suppression observed at SPS 
and RHIC must be due to the dissociation of the higher excited states 
\X~and \P. The suppression onset for these is predicted to lie around 
$\e \simeq 1$ GeV/fm$^3$, and once they are gone, only the unaffected  
$\j(1S)$ production remains. Hence the \J~survival probability (under
anomalous suppression) should be the same for central $Au-Au$ collisions 
at RHIC as for central $Pb-Pb$ collisions at the SPS.

\medskip

A further check to verify that the observed \J~production in central
collisions is indeed due to the unmodified survival of the directly
produced $1S$ state is provided by its transverse momentum behaviour.
Initial state parton scattering causes a broadening of the $p_T$
distributions of charmonia \cite{KP}-\cite{HS-91}: the gluon from the 
proton projectile in $pA$ collisions can scatter a number of times in 
the target nucleus before fusing with a target gluon to produce a $\C$. 
Assuming the protonic gluon to undergo a random walk through the target 
leads to
\be 
\langle p_T^2 \rangle_{pA} = \langle p_T^2 \rangle_{pp} + N_c^A \delta_0
\ee
for the average squared transverse momentum of the observed \J. Here
$N_c^A$ specifies the number of collisions of the gluon before the parton 
fusion to $\C$, and $\delta_0$ the kick it receives at each collision.
The collision number $N_c^A$ can be calculated in the Glauber formalism;
here $\sigma_{abs}$ has to be included to take into account the presence 
of cold nuclear matter, which through a reduction of \J~production shifts
the effective fusion point further ``down-stream'' \cite{KNS}.

\medskip

In $AA$ collisions, initial state parton scattering occurs in both
target and projectile, and the corresponding random walk form becomes
\be
\langle p_T^2 \rangle_{AA} = \langle p_T^2 \rangle_{pp} + N_c^{AA} \delta_0;
\ee
here $N_c^{AA}$ denotes the sum of the number of collisions in the 
target and in the projectile, prior to parton fusion.  It can again be
calculated in the Glauber scheme including $\sigma_{abs}$. The crucial
point now is that if the observed \J's in central $AA$ collisions are
due to undisturbed $1S$ production, then the centrality dependence of 
the $p_T$ broadening is fully predicted by such initial state parton 
scattering \cite{KKS}. In contrast, any onset of anomalous
suppression of the $\j(1S)$ would lead to a modification of the random
walk form \cite{KNS}.

\medskip

\begin{figure}[htb]
{\hskip1cm \epsfig{file=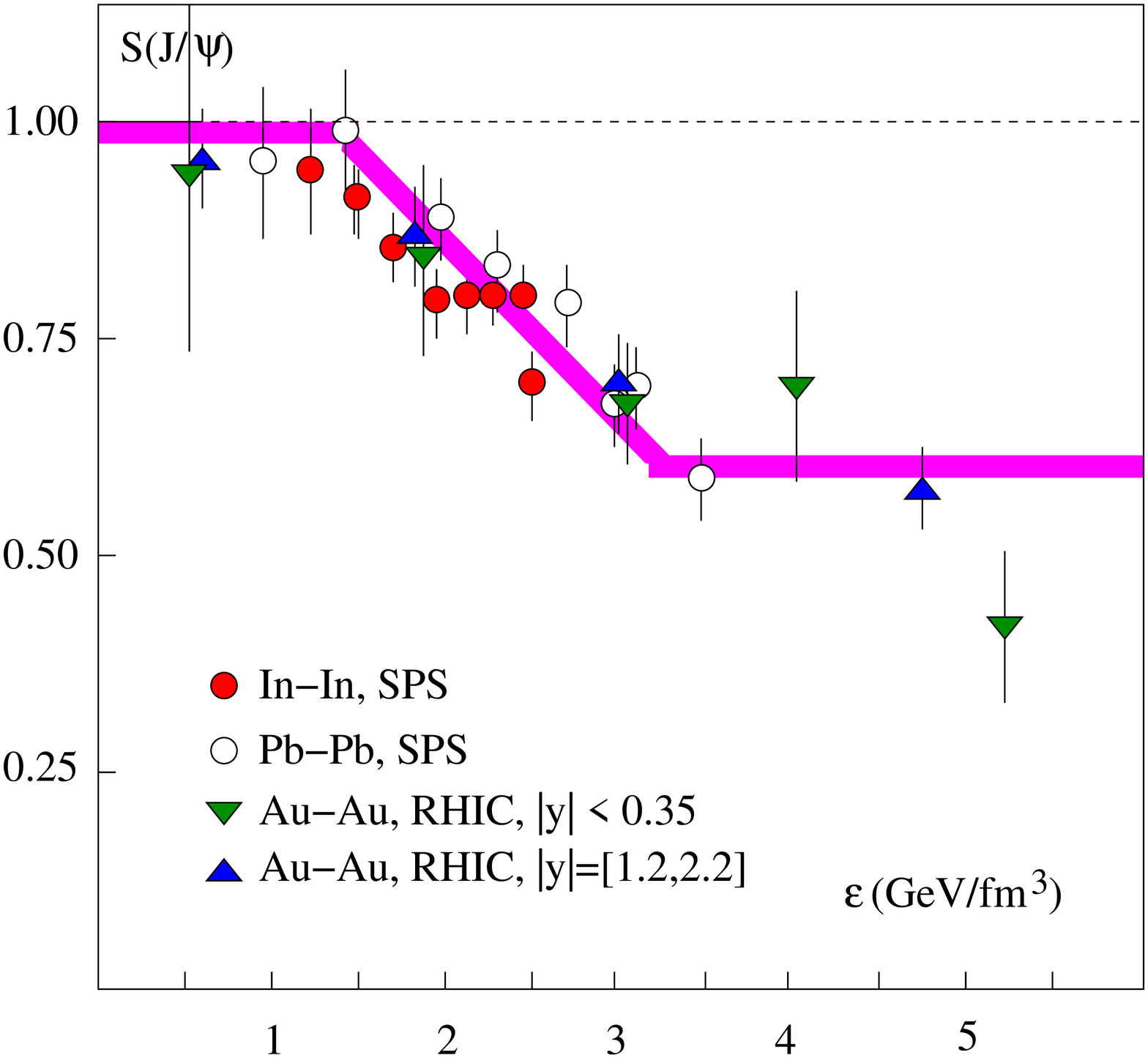,width=6cm,height=4.5cm} \hfill
\epsfig{file=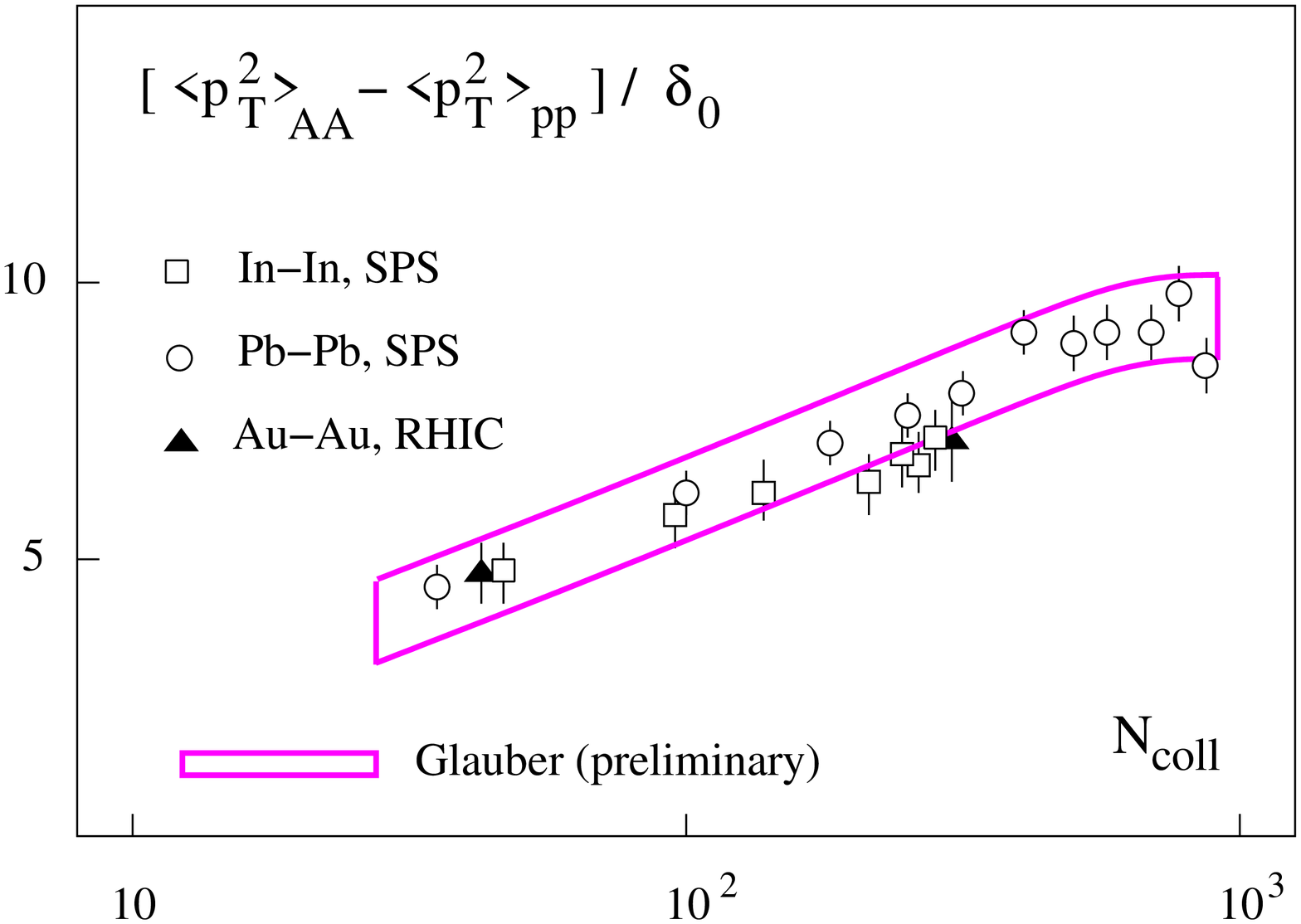,width=6cm,height=4.5cm} \hskip1cm}
\caption{\J~survival and transverse momentum at SPS and RHIC}
\label{data}
\end{figure}

In Fig.\ \ref{data} we summarize the predictions for \J~survival and 
transverse momentum behaviour in $AA$ collisions at SPS and RHIC, as 
they emerge from our present state of knowledge of statistical QCD. 
Included are some preliminary and some final data; for a discussion 
of the data analysis and selection, see ref.\ \cite{KKS}. 

\medskip

\medskip

We conclude that the present experimental results are compatible with
the present information from statistical QCD. This was not the case
previously, and such a conclusion can be drawn today because of 
several changes in our theoretical and experimental understanding: 
\begin{itemize}
\vspace*{-0.2cm}
\item{As already noted, statistical QCD presently puts the onset of direct 
\J~suppression at energy densities beyond the RHIC range; previous onset 
values were much lower (see, e.g., ref.\ \cite{DPS1}).}
\vspace*{-0.2cm}
\item{SPS $In-In$ data \cite{In}
suggest an onset of anomalous suppression at 
$\e \simeq$ 1 GeV/fm$^3$; previous onset values from $Pb-Pb$ and $S-U$
interactions were considerably higher, with $\e \simeq 2 - 2.5$ GeV/fm$^3$
(see, e.g., ref.\ \cite{Abreu02}).}
\vspace*{-0.2cm}
\item{within statistics, there is no further drop of the \J~survival rate 
below 50 - 60 \%, neither at RHIC nor at the SPS; a second drop in very
central SPS $Pb-Pb$ data (see, e.g., ref.\ \cite{Abreu02}) is no longer 
maintained.}
\vspace*{-0.2cm}
\end{itemize}

\vskip0.5cm

\noindent{\large \bf 4.\ \J~Enhancement by Regeneration}

\bigskip

In this section we want to consider the possibility that the medium
produced in high energy nuclear collisions differs from the deconfined
state of matter studied in finite temperature QCD. The basic idea here 
is that nuclear collisions initially produce more than the thermally
expected charm, and that this excess, if it survives, may lead to a new
form of combinatorial charmonium production at hadronization.

\medskip 

\begin{figure}[htb]
\centerline{\epsfig{file=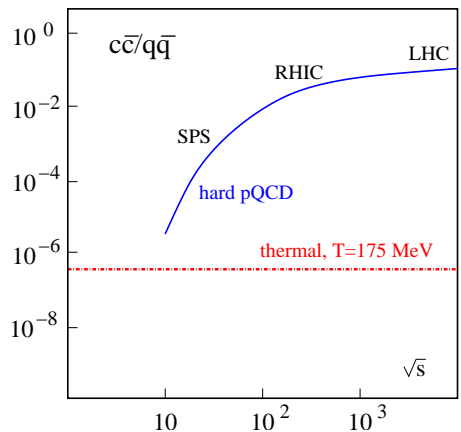,width=6cm}}
\caption{Thermal vs.\ hard charm production in $pp$ collisions}
\label{initial}
\end{figure}

A crucial aspect in the QGP argumentation of the previous sections was 
that charmonia, once dissociated, cannot be recreated at the hadronization 
stage, since the abundance of charm quarks in an equilibrium QGP is far 
too low to allow this. The thermal production rate for a $\C$ pair, relative 
to a pair of light quarks, is 
\be
{c\bar c /q \bar q} \simeq \exp\{-2m_c/T_c\} \simeq 3.5 \times 10^{-7},
\ee
with $m_c = 1.3$ GeV for the charm quark mass and $T_c=0.175$ GeV for the
transition temperature. The initial charm production in high energy
hadronic interactions, however, is a hard non-thermal process, and the
resulting rates from perturbative QCD are considerably larger. We 
illustrate this for $pp$ collisions in Fig.\ \ref{initial},
with $c\bar c/q\bar q = \sigma_{\C} / \sigma_{in}$ \cite{LW,PDG}.
Moreover, in $AA$ interactions
the resulting $c/\bar c$ production rates grow with the number $N_{coll}$ 
of nucleon-nucleon collisions, while the light quark production rate grows 
(at least in the present energy regime) essentially as the number 
$N_{part}$ of participant nucleons, i.e., much slower. At high
collision energies, the initial charm abundance in $AA$ collisions is 
thus very much higher than the thermal value. What happens to this 
excess in the course of the collision evolution?

\medskip

The basic assumption of the regeneration approach \cite{PBM} - \cite{Rapp}
is that the initial charm excess is maintained throughout the subsequent
evolution, i.e., that the initial chemical non-equilibrium will persist
up to the hadronization point. If that is the case, a $c$ from a given
nucleon-nucleon collision can at hadronization combine with a $\bar c$
from a different collision (``off-diagonal'' pairs) to create a \J.
This pairing provides a new {\sl exogamous} charmonium production mechanism, 
in which the $c$ and the $\bar c$ in a charmonium state have differen
parents, in contrast to the {\sl endogamous} production in a $pp$ collision.
At sufficiently high energy, this mechanism will lead to enhanced
\J~production in $AA$ collisions in comparison to the scaled $pp$ rates.
When should this enhancement set in?
 
\medskip

In present work \cite{PBM} - \cite{Rapp}, it is first assumed that
the direct \J~production is strongly suppressed for $\e \geq 3$ GeV/fm$^3$.
This is evidently in contrast to the statistical QCD results discussed in
the previous sections; however, we recall the caveat that the temperature 
dependence of the charmonium widths is so far not known. Moreover, it is
of course always possible that the medium produced in nuclear collisions
is quite different from the quark-gluon plasma of statistical QCD. 
Next, it is either assumed that the regeneration rate is determined by 
statistical combination in a QGP \cite{PBM} or a specific in-medium 
$\C$ recombination process is invoked, depending on the expansion geometry 
and the momentum distribution of the produced charm quarks \cite{Thews}.
In general, however, if a $c$ and a $\bar c$ meet under the right kinematic 
conditions, they are taken to form a \J. An evolution towards a QGP in 
{\sl chemical} equilibrium would also require annihilation at this point.

\medskip

\begin{figure}[htb]
\centerline{\epsfig{file=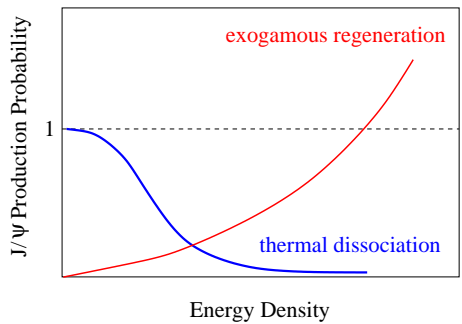,width=6cm}}
\caption{Regeneration of \J~production}
\label{recom}
\end{figure}

To account for the \J~production rates observed at RHIC, it is 
assumed that the new exogamous production just compensates the proposed
decrease of the direct endogamous $1S$ production, as illustrated in
Fig.\ \ref{recom}. At the LHC, with much higher energy densities, one 
should then observe a \J~enhancement relative to the rates expected
from scaled $pp$ results.

\vskip-0.2cm

We thus have to find ways of distinguishing between the two scenarios
discussed here: the sequential suppression predicted by an equilibrium 
QGP or a stronger direct suppression followed by a \J~regeneration in a
medium with excess charm. Fortunately the basic production patterns in
the two cases are very different, so that one may hope for an eventual
resolution.   

\medskip

\begin{figure}[htb]
\hfill{\epsfig{file=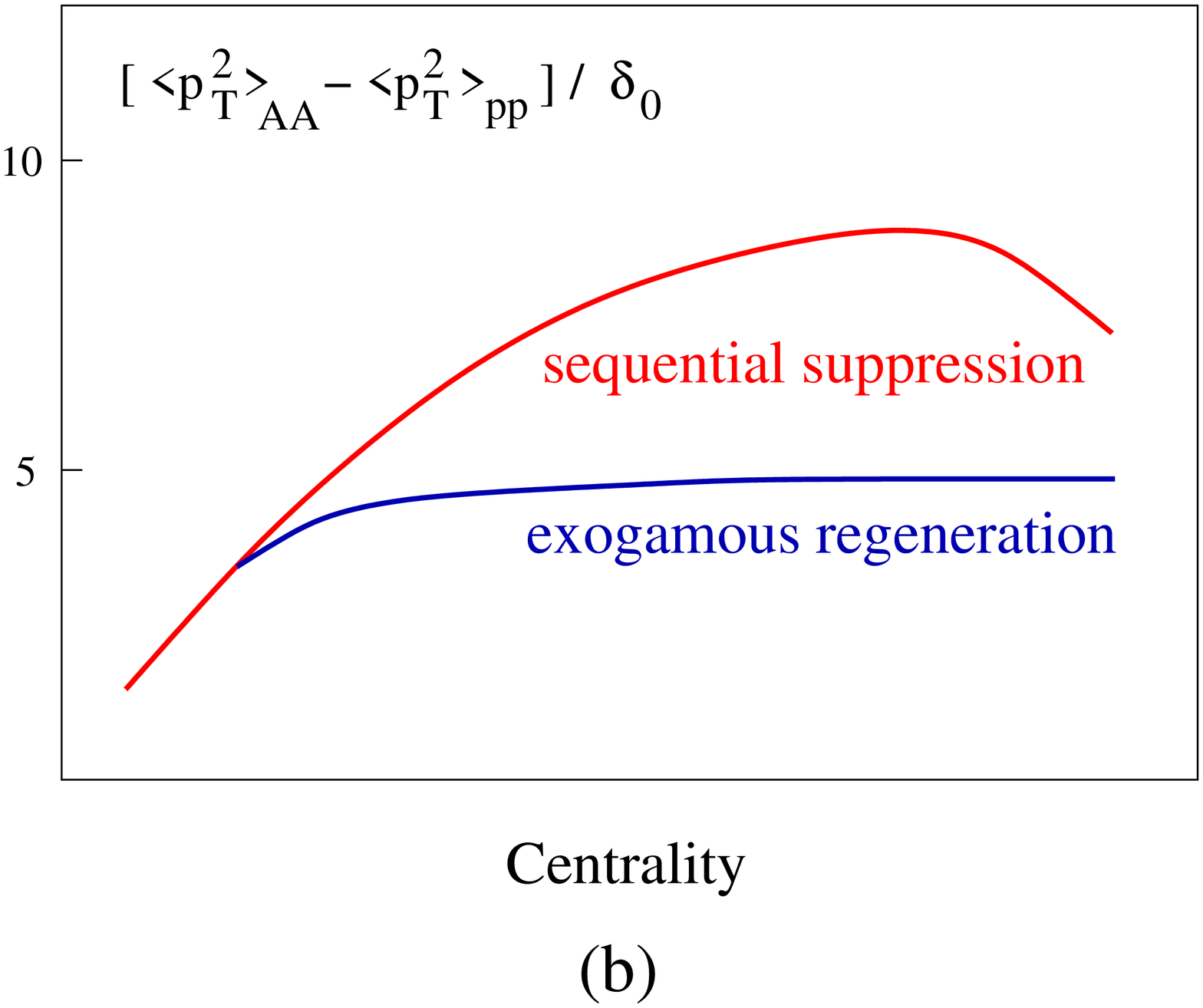,width=6cm,height=4.5cm} \hskip1cm}
\caption{Sequential suppression vs.\ regeneration: \J~survival (a) and 
$p_T$-behaviour (b)}
\label{comp}
\end{figure}

The overall \J~survival probability in the two cases is illustrated in
Fig.\ \ref{comp}a. Sequential suppression provides a step-wise reduction:
first the higher excited charmonium states are dissociated and thus their 
feed-down contribution disappears; at much higher temperature, the $\j(1S)$  
itself is suppressed. Both onsets are in principle predicted by lattice QCD
calculations. In the regeneration scenario, the thermal dissociation of 
all ``diagonal'' \J~production is obtained by extrapolating SPS data to
higher energy densities. The main prediction of the approach is therefore
the increase of \J~production with increasing energy density. Ideally, the
predictions for the LHC are opposite extremes \cite{KKS,Andron,Nu}, 
providing of course that here the feed-down from $B$-decay is properly 
accounted for.

\begin{figure}[htb]
{\hskip1cm \epsfig{file=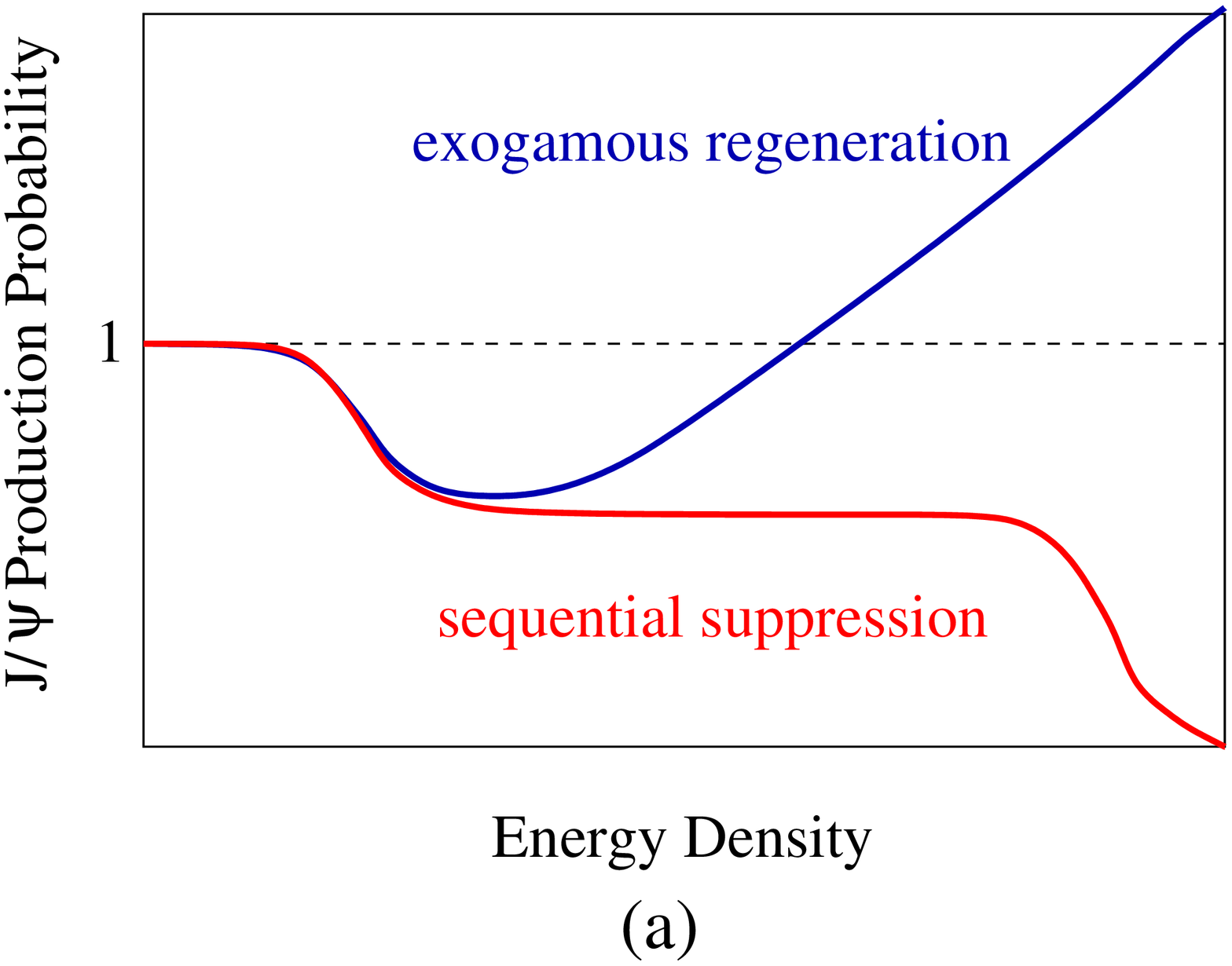,width=6cm,height=4.5cm}} 
\end{figure}

\vspace*{-5.45cm}

The expected transverse momentum behaviour in the two cases is also
quite different. In the region of full $\j(1S)$ survival, sequential
suppression predicts the normal random walk pattern specified through
$pA$ studies; the eventual dissociation of direct $\j(1S)$ states then
leads to an anomalous suppression also in the average $p_T^2$ \cite{KNS}.   
Regeneration alone basically removes the centrality dependence, since
the different partners come from different collisions. It is possible
to introduce some small centrality dependence \cite{MT}, but the random 
walk increase is essentially removed. The resulting behaviour is 
schematically illustrated in Fig.\ \ref{comp}b. -- More generally, the 
quarkonium momentum distributions, whether transverse or longitudinal, 
should in the regeneration scenario be simply a convolution of the 
corresponding open charm distributions; this provides a further 
check \cite{MT}. 

\vskip0.8cm

\noindent{\large \bf 5.\ Conclusions}

\bigskip

\begin{itemize}
\vspace*{-0.2cm}
\item{In statistical QCD, the spectral analysis of quarkonia provides  
a well-defined way to determine the temperature and energy density of the
QGP.}
\vspace*{-0.2cm}
\item{If nuclear collisions produce a quark-gluon plasma in equilibrium, 
the study of quarkonium production can provide a direct way to connect 
experiment and statistical QCD.}
\vspace*{-0.2cm}
\item{For a QGP with surviving charm excess, off-diagonal quarkonium 
formation by statistical combination may destroy this connection and
instead result in enhanced \J~production.} 
\end{itemize}

\vskip0.5cm

\centerline{\large \bf Acknowledgements}

\bigskip

It is a pleasure to thank R.\ Arnaldi, S.\ Digal, R.\ Granier de Cassagnac, 
F.\ Karsch, C.\ Louren{\c c}o, A.\ M\'ocsy, M.\ Nardi, P.\ Petreczky, 
R.\ L.\ Thews, R.\ Vogt and H.\ W\"ohri for helpful and stimulating 
discussions.

\end{document}